\documentclass[aps,pra,superscriptaddress,amsmath,amssymb,preprintnumbers,showpacs]{revtex4}

\usepackage{graphicx}   
\usepackage{amsmath}            

\begin{document}

\newcommand{\bra}[1]{\langle #1 \vert}
\newcommand{\ket}[1]{\vert #1 \rangle}
\newcommand{\scl}[2]{( #1 \vert #2 )}
\newcommand{\dert}[1]{\frac{d}{dt}#1}
\newcommand{\dertn}[2]{\frac{d^{#2}}{dt^{#2}}#1}
\newcommand{\derpt}[1]{\frac{\partial}{\partial t}#1}
\newcommand{\derpn}[3]{\frac{\partial^{#2}}{\partial {#3}^{#2}}#1}
\newcommand{\med}[1]{\langle#1\rangle}
\newcommand{\mte}[1]{\| #1 \|}
\newcommand{\abs}[1]{\vert #1 \vert}
\newcommand{\sa}[1]{{#1}_{\Sigma}}
\newcommand{\tr}[2]{tr_{#2}\big{\{}#1\big{\}}}
\newcommand{\Tr}[1]{tr \big{\{}#1\big{\}} }
\newcommand{\defi}[1]{\stackrel{def.}{=} #1 }


\title{The rotating wave system-reservoir coupling: limitations and meaning in the non-Markovian regime}



\author{F. Intravaia}
\affiliation{INFM and MIUR, Dipartimento di Scienze Fisiche ed
Astronomiche dell'Universit\`{a} di Palermo, via Archirafi 36,
90123 Palermo, Italy.}

\author{S. Maniscalco}
\email{sabrina@fisica.unipa.it}

\affiliation{INFM and MIUR, Dipartimento di Scienze Fisiche ed
Astronomiche dell'Universit\`{a} di Palermo, via Archirafi 36,
90123 Palermo, Italy.}

\author{A. Messina}

\affiliation{INFM and MIUR, Dipartimento di Scienze Fisiche ed
Astronomiche dell'Universit\`{a} di Palermo, via Archirafi 36,
90123 Palermo, Italy.}

\date{\today}

\begin{abstract}
This paper deals with the dissipative dynamics of a quantum
harmonic oscillator interacting with a bosonic reservoir. The
Master Equations based on the Rotating Wave and on the
Feynman-Vernon system--reservoir couplings are compared
highlighting differences and analogies. We discuss quantitatively
and qualitatively the conditions under which the counter rotating
terms  can be neglected. By comparing the analytic solution of the
heating function relative to the two different coupling models we
conclude that, even in the weak coupling limit, the counter
rotating terms give rise to a significant contribution in the
non--Markovian short time regime. The main result of this paper is
 that such a contribution is actually experimentally
measurable and thus relevant for a correct description of the
system dynamics.
\end{abstract}

\pacs{03.65.Yz,03.65.Ta}

\maketitle

\section{Introduction}
During the last few decades a huge deal of attention has been
devoted to the study of the quantum dynamics of  dissipative
systems. The theory of open quantum systems, indeed, is essential
for the understanding of a variety of physical phenomena in
different fields of physics, such as, for example, quantum optics
and solid state physics \cite{gardiner}. Moreover, very recently,
there has been an increasing interest in the effects of
decoherence, due to the unavoidable coupling with external
environment\cite{ptzurek,zurek81,zurek82}, on the dynamics of
quasi-closed systems used for quantum computing and quantum
information processing. The usual approach for studying
decoherence and dissipation effects starts by prescribing a total
Hamiltonian for the closed total system (system+reservoir). Then,
after tracing over the reservoir variables and performing, if
necessary, appropriate approximations, one finally derives a
Master Equation ruling the dynamics of the dissipative quantum
system. One of the most commonly done assumptions for describing
open quantum systems is the so-called Born-Markov approximation
which basically consists in neglecting memory effects of the
reservoir. In other words one assumes that the correlation time of
the reservoir, characterizing the time scale on which the
reservoir memory would feed back to the system, is much shorter
than the typical system time scale. When such condition is
satisfied it is possible to derive a Master Equation describing
the time evolution of the dissipative system for times longer than
the correlation time of the reservoir. Under this approximation
the resulting Master Equation is called Markovian Master Equations
and of course does not describe appropriately systems interacting
with natural or engineered structured reservoirs, such as atoms
decaying in photonic band gap materials or atom lasers. It has
been very recently demonstrated by Ahn {\it et al.} \cite{Ahn}
that non-Markovian reservoirs may be of potential interest for
quantum information processing since a quantum system is decohered
slower in a non-Markovian reservoir than in a Markovian one.

In this paper we firstly derive the time--convolutionless Master
Equation describing a quantum harmonic oscillator of frequency
$\omega_0$ interacting with a bosonic reservoir represented as an
infinite chain of harmonic oscillators of frequencies
$\{\omega_i\}$ \cite{fkm65,fk87,legget}. The method used,
exploiting a superoperatorial formalism, leads to a non-Markovian
Master Equation spoiled of reservoir memory
kernels\cite{tcl,petruccionebook}. In words one says that such a
Master equation is local in time. Our aim is to analyze
differences and analogies in the dynamical behaviour of this
specific open system in correspondence with two different prefixed
system-reservoir couplings. The first choise is the following:
\begin{equation}
\hat{H}^{\rm RWA}_{sr} = \alpha  \sum_{i=0}^{\infty}
\hbar\sqrt{\frac{\omega_i}{2}}\left( g_i \hat{a} \hat{b}_i^{\dag}
+ h.c. \right) \equiv \alpha \left(\hat{a} \hat{R}^{\dag} +
 \hat{a}^{\dag} \hat{R}\right), \label{eq:intro1}
\end{equation}
usually referred to as Rotating Wave (RW) coupling. In Eq.
(\ref{eq:intro1}), $\hat{a}$ and $\hat{b}_i$ are the annihilation
operators of the system and reservoir harmonic oscillators
respectively and $\alpha$ is  the adimensional coupling constant.
Note that, for the sake of simplicity, in the paper we use
adimensional position and momentum operators for the system
oscillator.


The second form of the system--reservoir interaction Hamitonian
 examined in this paper is the so-called Feynman-Vernon
(FV) coupling \cite{fvc}:
\begin{equation}
\hat{H}_{sr} = \alpha  \hat{X}\sum_{i=0}^{\infty}
\hbar\sqrt{\omega_i}\left( g^*_i\hat{b}_i + g_i\hat{b}_i^{\dag}
\right) \equiv \alpha\hat{X} \left(\hat{R} +
 \hat{R}^{\dag}\right), \label{eq:intro2}
\end{equation}
where the operator $\hat{X}$ is related to the creation and
annihilation operators of the quantum harmonic oscillator simply
as
\begin{equation}
\hat{X}=\frac{1}{\sqrt{2}}\left( \hat{a} + \hat{a}^{\dag} \right).
\label{eq:intro3}
\end{equation}

While the first interaction Hamiltonian is very often used in
describing quantum optics systems \cite{cfunction} and atom lasers
\cite{atomlaser}, the second one leads to the Master Equation for
Quantum Brownian Motion \cite{bmotion}. Using the Hamiltonian
given by Eq. (\ref{eq:intro1}) instead of the more general one
given by Eq. (\ref{eq:intro2}) is usually motivated saying that
the counter rotating terms $ \hat{a} \hat{b}_j $ and
$\hat{a}^{\dag} \hat{b}_j^{\dag}$, appearing in Eq.
(\ref{eq:intro2}), do not conserve the total unperturbed energy
and thus give a negligible contribution to the system dynamics in
the weak coupling limit \cite{cfunction}.

The main result of this paper is that, in the non-Markovian
regime, the contribution given by the counter-rotating terms  is
not negligible and experimentally measurable, also when the weak
coupling limit is invoked.

The paper is structured as follows. In Sec. II we introduce the
superoperator formalism for the derivation of non-Markovian
generalized Master Equations. In Sec. III we specialize the
generalized Master Equations to the cases of Rotating Wave and
Feynman-Vernon couplings and we compare them in Sec. IV. Finally
in Sec. V we present conclusions.

\section{Derivation of the Master Equation: an operatorial approach}
Let us consider an open quantum system interacting with an
environment whose physical nature needs not to be specified at
this moment. We indicate the total Hamiltonian as follows
\begin{equation}
\hat{H} = \hat{H}_0 + \hat{H}_E + \alpha \hat{H}_{\rm int},
\label{eq:secI1}
\end{equation}
where $\hat{H}_0$, $\hat{H}_E$ and $\hat{H}_{\rm int}$  stand for
the system, environment and interaction Hamiltonians respectively
and $\alpha$ is the coupling constant. The Van Neumann-Liouville
equation for the total system, in the interaction picture, is the
following
\begin{equation}
\frac{d \tilde{\rho}(t)}{dt} = \frac{\alpha}{i \hbar}
\left[\hat{H}_{I}(t), \tilde{\rho}(t)\right] \equiv
\frac{\alpha}{i \hbar} \mathbf{H}_I^{S} (t) \tilde{\rho}(t).
\label{eq:secI2}
\end{equation}
In Eq. (\ref{eq:secI2}), $\tilde{\rho}$ and $\hat{H}_I(t)$ are the
density matrix and the interaction Hamiltonian of the total system
respectively, in the interaction picture, and the superoperator
$\mathbf{H}_I^{S} (t)$ is defined as $\mathbf{H}_I^{S} (t) =
[\hat{H}_I(t), \cdot \;]$. In the rest of the paper, given a
certain operator $\hat{A}$,  we will use the following notation
for the \lq\lq commutator \rq\rq and \lq\lq anticommutator\rq\rq
superoperators:
\begin{equation}
\mathbf{A}^{S} = [\hat{A}, \cdot \; ] \hspace{1cm}
\mathbf{A}^{\Sigma} = \{\hat{A}, \cdot\;  \}. \label{eq:secI3}
\end{equation}
In deriving the generalized Master Equation we assume that at
$t=0$ system and environment are uncorrelated, that is
$\tilde{\rho}(0)= \hat{\rho} (0) \bigotimes \hat{\rho}_E(0)$, with
$\hat{\rho}$ and $ \hat{\rho}_E$ density matrices of system and
environment respectively and that the environment is stationary,
i.e. $[\hat{H}_E,\hat{\rho}_E]=0$.

A formal solution of Eq. (\ref{eq:secI2}) can be written as
\begin{equation}
\tilde{\rho}(t) = \mathbf{T}(t)  \tilde{\rho}(0), \label{eq:secI4}
\end{equation}
where the superoperator $ \mathbf{T}(t)$ is defined as the
solution of the equation:
\begin{equation}
\dot{\mathbf{T}}(t) = \frac{\alpha}{i \hbar} \mathbf{H}_I^{S} (t)
\mathbf{T} (t), \label{eq:secI5}
\end{equation}
with $\mathbf{T} (0) = \mathbf{1}$. Remembering that
$\hat{\rho}(t)= Tr_{\rm E} \left\{ \tilde{\rho}(t)\right\}$ and
that $\tilde{\rho}(0) =\hat{\rho}(0) \bigotimes \hat{\rho}_E(0)$,
after tracing over the environmental variables, Eq.
(\ref{eq:secI4}) becomes
\begin{eqnarray}
\hat{\rho}(t) &=& Tr_{\rm E} \left\{ \mathbf{T} (t)
\hat{\rho}_E(0)
\right\} \hat{\rho}(0) \nonumber \\
&\equiv& \langle \mathbf{T} (t)\rangle \hat{\rho}(0) \equiv(1+
\mathbf{M} (t)) \hat{\rho}(0) , \label{eq:secI5a}
\end{eqnarray}
where we have indicated with $\langle  \mathbf{T} (t)\rangle $ the
superoperator $(1+ \mathbf{M} (t))= Tr_{\rm E} \left\{ \mathbf{T}
(t) \hat{\rho}_E(0) \right\}$, acting on the space
$\mathcal{H}_{\rm s} \bigotimes \mathcal{H}_{\rm s}^*$, with
$\mathcal{H}_{\rm s}$ Hilbert space of the system. Differentiating
now Eq. (\ref{eq:secI5a}) yields
\begin{equation}
\frac{d \hat{\rho}(t)}{dt} =  \dot{\mathbf{M}} (t) \hat{\rho}(0).
\label{eq:secI6}
\end{equation}
Inserting in Eq.(\ref{eq:secI6}) the expression for
$\hat{\rho}(0)$ obtained inverting Eq. (\ref{eq:secI5a}) gives
\begin{equation}
\frac{d \hat{\rho}(t)}{dt} = \left( \dot{\mathbf{M}}(t) \right)
\left[1+ \mathbf{M} (t)\right]^{-1} \hat{\rho}(t) \equiv
\mathbf{K} (t) \hat{\rho}(t). \label{eq:secI7}
\end{equation}
In the previous equation we have defined a new superoperator
$\mathbf{K} (t)$ acting on the space $\mathcal{H}_{\rm s}
\bigotimes \mathcal{H}_{\rm s}^*$ too. At this point it is worth
spending few words on the existence of $\mathbf{K} (t)$, that is
of the inverse superoperator $\left[1+ \mathbf{M}
(t)\right]^{-1}$. To this aim, we recast $\mathbf{K} (t)$ in the
form
\begin{equation}
\mathbf{K} (t) = \left( \dot{\mathbf{M}}(t) \right) \left[1+
\mathbf{M} (t)\right]^{-1}=\left( \dot{\mathbf{M}}(t)
\right)\sum_n \left(-\mathbf{M} (t)\right)^n \label{eq:secI8}
\end{equation}
As discussed in \cite{petruccione} for weak couplings such series
converges at any  time $t$. For generic coupling, however, the
convergence radius of the series depends both on $\alpha$ and on
$t$. For this reason one has always to pay special attention to
such a problem when working for intermediate or even strong
coupling regimes.

Now, it is easy to convince oneself that a formal solution of
Eq.(\ref{eq:secI5}) may be written as
\begin{eqnarray}
\mathbf{T} (t) = \exp_c \left[ \frac{\alpha}{i \hbar} \int_0^t
\mathbf{H}_I^{S} (t_1)\:dt_1 \right] \label{eq:secI9a} \equiv
\sum_{n=0}^{\infty} \left(\frac{\alpha}{i \hbar}\right)^n
 \int_0^t \int_0^{t_1} \cdot\cdot\cdot \int_0^{t_{n-1}}
\mathbf{H}_I^{S} (t)\mathbf{H}_I^{S} (t_1) \cdot\cdot\cdot
\mathbf{H}_I^{S} (t_n) \; dt_1 \cdot\cdot\cdot dt_n,
\label{eq:secI9b}
\end{eqnarray}
where the subscript $c$ in the exponential stands for the Dyson
chronological order, i.e. $t_n>t_{n-1} \cdot\cdot\cdot
>t_1>t$. Inserting such an expression into Eq. \eqref{eq:secI7} with the help of Eq. \eqref{eq:secI5a} and
collecting all the terms proportional to the same power in
$\alpha$, it is possible to demonstrate that the following
expansion holds:
\begin{equation}
\mathbf{K} (t) = \sum_{n=0}^{\infty} \mathbf{k}_n (t)
\label{eq:secI11}
\end{equation}
with
\begin{equation}
\mathbf{k}_n (t) = \left(\frac{\alpha}{i \hbar}\right)^n \int_0^t
\int_0^{t_1} \cdot\cdot\cdot \int_0^{t_{n-1}} \langle\langle
\mathbf{H}_I^{S} (t)\mathbf{H}_I^{S} (t_1) \cdot\cdot\cdot
\mathbf{H}_I^{S} (t_n) \rangle \rangle_{o.c.} \; dt_1
\cdot\cdot\cdot dt_n. \label{eq:secI12}
\end{equation}
In the previous equation we have indicated with $\langle \langle
\cdot\cdot\cdot \rangle \rangle_{o.c.}$ the temporal ordered
cumulants \cite{kubo}. As an example, we report the expression of
the first and second cumulants, respectively:
\begin{eqnarray}
\langle \langle \mathbf{H}_I^{S}(t) \rangle \rangle_{o.c.} &=&
\langle
\mathbf{H}_I^{S}(t) \rangle \nonumber \\
\langle \langle \mathbf{H}_I^{S}(t) \mathbf{H}_I^{S}(t_1)\rangle
\rangle_{o.c.} &=& \langle \mathbf{H}_I^{S}(t)
\mathbf{H}_I^{S}(t_1) \rangle -\langle \mathbf{H}_I^{S}(t) \rangle
\langle
 \mathbf{H}_I^{S}(t_1) \rangle.
\end{eqnarray}

The form of Eq. \eqref{eq:secI12} resembles the result obtained by
Van Kampen in the context of stochastic differential equations
\cite{artvk1,artvk2}.

The origin of the expansion given  by Eqs. \eqref{eq:secI11} and
\eqref{eq:secI12} can be understood as follows. Let us write the
superoperator $\mathbf{K} (t)$ defined by  Eq. \eqref{eq:secI8} in
following symbolic form
\begin{equation}
\mathbf{K} (t) = \frac{\delta }{\delta t} \; ln \left[\langle
\exp_c \left( \alpha \int_0^t \mathbf{H}_I^{S} (t_1) \:dt_1
\right) \rangle \right], \label{eq:secI10}
\end{equation}
where
\begin{equation}
\frac{\delta }{\delta t} \; F[\mathbf{A}(t)]\equiv
\left(\frac{d}{dt}\mathbf{A}(t)\right)F[\mathbf{A}(t)]
\end{equation}

In Eq. \eqref{eq:secI10} the symbol $\langle \cdot\cdot\cdot
\rangle=\tr{\cdots \hat{\rho}_E}{E}$ describes an operation of
average over the environmental degrees of freedom. The expression
$\langle \exp_c \left( \alpha\int_0^t \mathbf{H}_I^{S} (t_1)
\:dt_1) \right) \rangle $ can thus be seen as the generalization,
in the superoperator formalism, of the concept of characteristic
functional \cite{vkbook}. As a consequence the superoperator $ln
\left[\langle \exp_c \left( \alpha \int_0^t \mathbf{H}_I^{S} (t)
\right) \rangle \right]$ is the generalization of the generator of
cumulants introduced in standard textbooks. This circumstance
makes it clear why the integrand in Eq. \eqref{eq:secI12} is
called temporal ordered cumulant. In view of Eq. \eqref{eq:secI10}
the existence problem of the superoperator $\mathbf{K}(t)$ can be
traced back to the convergence of the series of cumulants in
Eq.(\ref{eq:secI11}).

In order to derive the explicit form of the generalized Master
Equation, let us assume a bilinear interaction Hamiltonian of the
form:
\begin{equation}
\hat{H}_I (t)= \alpha \sum_{i=1}^n \hat{A}_i(t)  \hat{E}_i(t)=
\alpha \hat{\vec{A}}(t) \cdot  \hat{\vec{E}}(t), \label{eq:secI13}
\end{equation}
where $\hat{\vec{A}}(t)\equiv
\left(\hat{A}_1(t),\hat{A}_2(t),\dots,\hat{A}_n(t) \right)$ and
$\hat{\vec{E}}(t)\equiv
\left(\hat{E}_1(t),\hat{E}_2(t),\dots,\hat{E}_n(t) \right)$ are
system and environment operators respectively. In the weak
coupling limit we may stop the cumulant expansion given in
Eq.\eqref{eq:secI11} to the second order in the coupling constant.
In view of Eqs.\eqref{eq:secI7} and \eqref{eq:secI11}
\eqref{eq:secI12} , this leads to the following Master Equation
\begin{eqnarray}
\frac{d \hat{\rho}(t)}{dt} &=& \frac{\alpha}{i \hbar} \langle
\mathbf{H}_I^{S}(t) \rangle \hat{\rho}(t) \nonumber \\
&-& \frac{\alpha^2}{ \hbar^2} \int_0^t \left[ \langle
\mathbf{H}_I^{S}(t) \mathbf{H}_I^{S}(t_1) \rangle -\langle
\mathbf{H}_I^{S}(t) \rangle \langle
 \mathbf{H}_I^{S}(t_1) \rangle \right] dt_1 \;\hat{\rho}(t). \label{eq:secI14}
\end{eqnarray}
Assuming for simplicity that the form of the environmental density
matrix satisfy the condition $\med{\hat{E}}=0$ (as for example in
the case of a thermal reservoir), one can show that the first term
of Eq.(\ref{eq:secI14}) vanishes at every time $t$. The explicit
manipulation of the second term is presented in Appendix A and
leads to the following final form of the non-Markovian generalized
Master Equation:
\begin{eqnarray}\label{eq:secI15}
\frac{d \hat{\rho}(t)}{dt} &=& - \sum_{i,\:j=1}^n\left(\int_0^t
\left[ \kappa_{i,j} (\tau) \mathbf{A}_i^S(t)
\mathbf{A}_j^S(t-\tau) - i \mu_{i,j}(\tau) \mathbf{A}_i^S(t)
\mathbf{A}_j^{\Sigma}(t-\tau)
\right] d\tau \right) \hat{\rho}(t) \nonumber\\
&\equiv&-\left[\mathbf{D}(t)-\imath\mathbf{G}(t)\right]\hat{\rho}(t)
\equiv \mathbf{L}(t)\hat{\rho}(t), \label{Lt}
\end{eqnarray}
where definitions of Eq.\eqref{eq:secI3} have been used. In
Eq.\eqref{eq:secI15} we have introduced the environment
correlation $\kappa_{i,j}(\tau)$ and susceptibility
$\mu_{i,j}(\tau)$ matrices, with $\tau = t-t_1 $. Such quantities,
characterizing the temporal behavior of the environment, are
defined as follows
\begin{equation}
\kappa_{i,j}(\tau)= \frac{\alpha^2}{2 \hbar^2} \langle \{
\hat{E}_i(\tau),\hat{E}_j(0)\} \rangle, \label{eq:secI16a}
\end{equation}
\begin{equation}
\mu_{i,j}(\tau)= \frac{i\alpha^2}{2 \hbar^2} \langle [
\hat{E}_i(\tau),\hat{E}_j(0)] \rangle. \label{eq:secI16b}
\end{equation}

The form of Eq. \eqref{eq:secI15} has a clear physical meaning.
One can show that the superoperator $\mathbf{D}(t)$ is strictly
connected with diffusion (decoherence) processes only
\cite{leshouches}. The superoperator $\mathbf{G}(t)$, describing
the dissipation processes and frequency renormalization, on the
other hand, arises from a quantum mechanical treatment of the
environment and, indeed, vanishes when a semiclassical description
of the environment is used (see also Eq.\eqref{eq:secI16b})
\cite{leshouches}.


In the next section we further carry on the calculations in order
to obtain and compare the two non-Markovian Master Equations
corresponding to the Rotating Wave and Feynman-Vernon couplings
respectively.

\section{Rotating Wave and Feynman-Vernon couplings in the non-Markovian regime}

Let us consider a quantum harmonic oscillator whose Hamiltonian is
given by:
\begin{equation}
\hat{H}_0 =\frac{\hbar
\omega_0}{2}\left(\hat{P}^2+\hat{X}^2\right)=\hbar \omega_0 \left(
\hat{a}^{\dag} \hat{a} + 1/2 \right), \label{eq:secIII1}
\end{equation}
with $\omega_0$ frequency of the harmonic oscillator. The system
interacts with a bosonic reservoir at $T$ temperature of
Hamiltonian
\begin{equation}
\hat{H}_E = \hbar \sum_{i=0}^{\infty} \omega_i  \left(
\hat{b}^{\dag}_i \hat{b}_i + 1/2 \right), \label{eq:secIII2}
\end{equation}
with $\omega_i$ frequencies of the reservoir oscillators.

\subsection{Feynman-Vernon coupling}

Let us begin discussing the Feynman-Vernon interaction
Hamiltonian, given by Eq.(\ref{eq:intro2}). Such a coupling is of
the form of Eq.(\ref{eq:secI13}) where, in the interaction
picture, $\hat{A}(t) = \hat{X}(t)$ and $\hat{E}(t) = \hat{R}(t)+
\hat{R}^{\dag}(t)$. Our aim is to manipulate Eq.(\ref{eq:secI15})
in order to obtain the specific non-Markovian generalized Master
Equation appropriate for our system. In this section we will
sketch the main steps of the derivation. More details can be found
in  Appendix B.

First of all let us  write the Master Equation given in
Eq.(\ref{eq:secI15}) in the Schr\"{o}dinger picture. Introducing
the superoperator
\begin{equation}
\mathbf{T}_0 (t) = \exp \left[ \frac{1}{i \hbar} \mathbf{H}_0^{S}
t \right], \label{eq:secIII3}
\end{equation}
with $\hat{H}_0$ given by Eq.(\ref{eq:secIII1}), and transforming
in the Schr\"{o}dinger picture the superoperator $\mathbf{L}(t)$
defined in Eq.(\ref{Lt})
\begin{equation}
\mathbf{L}_S(t) = \mathbf{T}_0 (t) \; \mathbf{L}(t) \;
\mathbf{T}_0^{-1} (t), \label{eq:secIII4a}
\end{equation}
our generalized Master Equation becomes
\begin{equation}
\frac{d \hat{\rho}_S(t)}{dt} =  \left[\frac{1}{i \hbar}
\mathbf{H}_{0}^S - \mathbf{D}_S(t) + i
\mathbf{G}_S(t)\right]\hat{\rho}_S(t), \label{eq:secIII5}
\end{equation}
with $\hat{\rho}_S$ density matrix of the harmonic oscillator in
the Schr\"{o}dinger picture. In Appendix B we show that the
superoperators $\mathbf{D}_S(t)$ and $\mathbf{G}_S(t)$ can be
recast in the form
\begin{subequations}
\begin{equation}
\mathbf{D}_S(t) =\int_0^t \kappa (\tau) \mathbf{X}^S \left(
\cos{\omega_0 \tau } \mathbf{X}^S - \sin{\omega_0 \tau}
\mathbf{P}^S \right)d\tau, \label{eq:secIII6a}
\end{equation}
\begin{equation}
\mathbf{G}_S(t)= \int_0^t \mu (\tau) \mathbf{X}^{S} \left(
\cos{\omega_0 \tau } \mathbf{X}^{\Sigma} - \sin{\omega_0 \tau }
\mathbf{P}^{\Sigma} \right)d\tau, \label{eq:secIII6b}
\end{equation}
\end{subequations}
where $\mathbf{P}^S$ and $\mathbf{P}^{\Sigma}$ are the \lq\lq
commutator\rq\rq and \lq\lq anticommutator\rq\rq superoperators
associated to the operator
\begin{equation}
\hat{P}=  \frac{i}{\sqrt{2}}\left( \hat{a}^{\dag} - \hat{a}
\right) \label{eq:secIII7},
\end{equation}
and $\kappa(\tau)\equiv \kappa_{1,1 }(\tau)$ and $\mu(\tau) \equiv
\mu_{1,1} (\tau) $ are defined by Eqs.
(\ref{eq:secI16a})-(\ref{eq:secI16b}). Inserting
Eqs.(\ref{eq:secIII6a})-(\ref{eq:secIII6b}) into
Eq.(\ref{eq:secIII5}) we get
\begin{equation}
\frac{d \hat{\rho}_S(t)}{dt} = \frac{1}{i \hbar} \mathbf{H}_{0}^S
 -  \left[ \bar{\Delta}(t) (\mathbf{X}^S)^2 - \Pi(t) \mathbf{X}^S
\mathbf{P}^S - \frac{i}{2} r(t) (\mathbf{X^2})^S + i \gamma(t)
\mathbf{X}^S \mathbf{P}^{\Sigma} \right]\hat{\rho}_S(t).
\label{eq:secIII8}
\end{equation}
The time dependent coefficients appearing in the previous equation
are defined as follows
\begin{eqnarray}
\bar{\Delta}(t)&=& \int_0^t \kappa(\tau) \cos (\omega_0 \tau) d
\tau,
\label{eq:secIII9a} \\
\gamma(t) &=& \int_0^t \mu(\tau) \sin (\omega_0
\tau) d \tau, \label{eq:secIII9b} \\
\Pi(t) &=& \int_0^t \kappa(\tau) \sin (\omega_0
\tau) d \tau, \label{eq:secIII9c} \\
r(t)&=& 2 \int_0^t \mu(\tau) \cos (\omega_0 \tau) d \tau
\label{eq:secIII9d}.
\end{eqnarray}

From the form of Eq.(\ref{eq:secIII8}), and remembering that
$\mathbf{H}_{0}^S$ may be written as
\begin{equation}
\mathbf{H}_{0}^S = \frac{1}{2} \left[(\mathbf{P^2})^S +
(\mathbf{X^2})^S \right], \label{eq:secIII10}
\end{equation}
it is not difficult to convince oneself that the term having
coefficient $r(t)$ gives a renormalization of the oscillator
frequency.\\
As usually done in standard textbooks \cite{cfunction}, this term
can be included in the definition of $\omega_0$. In the following
we  neglect such term since it is possible to prove that such an
approximation is always justified in the weak coupling regime
$\alpha \ll 1$, provided that the reservoir frequency cut--off
remains finite.

Under these conditions, the Master Equation, in the interaction
picture with respect to $\tilde{H}_0$, assumes the form
\begin{equation}
\frac{d \hat{\rho}(t)}{dt} =  -  \left[ \bar{\Delta}(t)
(\mathbf{X}^S)^2 (t) - \Pi(t) \mathbf{X}^S (t) \mathbf{P}^S (t) +
i \gamma(t) \mathbf{X}^S (t) \mathbf{P}^{\Sigma} (t)
\right]\hat{\rho}(t). \label{eq:secIII11}
\end{equation}
The time dependent superoperators appearing in
Eq.(\ref{eq:secIII11}) are those related to the operators
\begin{eqnarray}
\hat{X}(t) &=& \hat{X} \cos(\omega_0 t) + \hat{P}
\sin(\omega_0 t), \label{eq:secIII12a} \\
\hat{P}(t) &=& \hat{P} \cos(\omega_0 t) -\hat{X} \sin(\omega_0 t).
\label{eq:secIII12b}
\end{eqnarray}
Eq. \eqref{eq:secIII11} can be exactly solved in an operatorial
way and the solution has an operatorial form \cite{prl}. This fact
may be exploited to fully disclose both the short time
non-Markovian and the asymptotic Markovian behaviors
characterizing the dynamics of the system, as we will see in
Section IV.

\subsection{Rotating Wave coupling}

The generalized Master Equation correspondent to the interaction
Hamintonian given by Eq. \eqref{eq:intro1}, derived following the
same procedure presented in the previous subsection, is (see also
Appendix C)

\begin{equation}
\frac{d \hat{\rho}(t)}{dt} = \left[- \bar{\Delta}^{\rm RWA}(t)
\mathbf{a^{\dag}}^S \mathbf{a}^S -\frac{\gamma^{\rm
RWA}(t)}{2}\left(\mathbf{a^{\dag}}^S
\mathbf{a}^{\Sigma}-\mathbf{a}^S
\mathbf{a^{\dag}}^{\Sigma}\right)+\frac{\imath}{2}r^{\rm
RWA}(t)\left(\mathbf{a^{\dag}}^S \mathbf{a}^{\Sigma}+\mathbf{a}^S
\mathbf{a^{\dag}}^{\Sigma}\right)\right]\hat{\rho}(t).\label{eq:secIV1}
\end{equation}

The time dependent coefficients appearing in this equation are
defined as follows
\begin{eqnarray}
\bar{\Delta}^{\rm RWA}(t)&=& \int_0^t \kappa^{\rm RWA}(\tau) d
\tau ,
\label{eq:secIV2a} \\
\gamma^{\rm RWA}(t) &=& \int_0^t \mu_R^{\rm RWA}(\tau)  d \tau , \label{eq:secIV2b} \\
r^{\rm RWA}(t) &=& 2 \int_0^t \mu_I^{\rm RWA}(\tau)  d \tau
\label{eq:secIV2c},
\end{eqnarray}
where
\begin{subequations}
\begin{equation}\label{coRWa}
\kappa^{\rm RWA}(\tau)=
\frac{\alpha^2}{2\hbar^2}\langle\{\hat{\tilde{R}}(\tau),\hat{\tilde{R}}^{\dag}(0)
\}+\{\hat{\tilde{R}}^{\dag}(\tau),\hat{\tilde{R}}(0) \} \rangle ,
\end{equation}
\begin{equation}\label{sureRWA}
\mu_R^{\rm RWA}(\tau)=
\frac{\imath\alpha^2}{2\hbar^2}\langle[\hat{\tilde{R}}(\tau),\hat{\tilde{R}}^{\dag}(0)
]+[\hat{\tilde{R}}^{\dag}(\tau),\hat{\tilde{R}}(0) ] \rangle ,
\end{equation}
\begin{equation}\label{suimRWA}
 \mu_I^{\rm RWA}(\tau)=
\frac{\imath\alpha^2}{2\hbar^2}\langle[\hat{\tilde{R}}(\tau),\hat{\tilde{R}}^{\dag}(0)
]-[\hat{\tilde{R}}^{\dag}(\tau),\hat{\tilde{R}}(0) ] \rangle ,
\end{equation}
\end{subequations}
and $\hat{\tilde{R}}(t)$ is the operator (see Appendix B)
\begin{equation}
\hat{\tilde{R}}(t)\equiv  \sum_{i=0}^{\infty}
\hbar\sqrt{\frac{\omega_i}{2}} g_i \hat{b}_i e^{-\imath(
\omega_i-\omega_0) t}.
\end{equation}

In the Schr\"{o}dinger picture Eq. \eqref{eq:secIV1} takes the
form
\begin{equation}
\frac{d \hat{\rho}_S(t)}{dt} = \left[\frac{1}{\imath
\hbar}\mathbf{H_0}^{S}+\imath r^{\rm
RWA}(t)\left(\mathbf{a^{\dag}a}\right)^S - \bar{\Delta}^{\rm
RWA}(t) \mathbf{a^{\dag}}^S \mathbf{a}^S -\frac{\bar{\gamma}^{\rm
RWA}(t)}{2}\left(\mathbf{a^{\dag}}^S
\mathbf{a}^{\Sigma}-\mathbf{a}^S
\mathbf{a^{\dag}}^{\Sigma}\right)\right]
\hat{\rho}_S(t)\label{Mes}
\end{equation}

The term proportional to $r^{\rm RWA}(t)$ gives rise to a
 renormalization of the oscillator frequency as for the
Feynman-Vernon case, described in the previous subsection.
Therefore, proceeding with the same considerations and passing to
the interaction picture, we obtain the following generalized
Master Equation for the system
\begin{eqnarray}
\frac{ d \hat{\rho}}{\partial t}= &-& \frac{\bar{\Delta}^{\rm
RWA}(t) + \gamma^{\rm RWA}(t)}{2} \left[ \hat{a}^{\dag} \hat{a}
\hat{\rho} - 2 \hat{a} \hat{\rho} \hat{a}^{\dag} + \hat{\rho}
\hat{a}^{\dag} \hat{a} \right]
\nonumber \\
&-& \frac{\bar{\Delta}^{\rm RWA}(t) - \gamma^{\rm RWA} (t)}{2}
\left[ \hat{a} \hat{a}^{\dag} \hat{\rho} - 2 \hat{a}^{\dag}
\hat{\rho} \hat{a} + \hat{\rho} \hat{a} \hat{a}^{\dag}
 \right]. \label{MeRWA}
\end{eqnarray}
Let us note that this Master Equation, differently from the one
obtained for the FV coupling (see Eq. \eqref{eq:secIII11}), is in
the Lindblad form as far as the time dependent coefficients
$\bar{\Delta}^{\rm RWA}(t) \pm \gamma^{\rm RWA} (t)$ are positive.
This is usually the case for typical reservoir spectra and
parameters, as we have discussed in \cite{pra}.

\section{Comparison between the RW and the Feynman-Vernon coupling models}

In the previous section we have seen that starting from a FV
coupling or from a RW coupling of an harmonic oscillator with a
thermal reservoir it is possible to obtain a generalized Master
Equation local in time describing the dynamics of the oscillator.
This fact is not surprising. Indeed, as underlined by Paz and
Zurek in \cite{leshouches}, \lq\lq perturbative Master Equations
can always be shown to be local in time \rq\rq. It is worth noting
that, as far as the FV interaction model is concerned, an exact
Master Equation, valid for every value of the coupling strength,
has been derived \cite{MEexact}.

The Master Equations we have derived in the paper are based on the
weak coupling assumption but do not rely on the Born-Markov
approximation so we are able to examine the non-Markovian short
time behavior of the system under study. In addition such
equations of course describe the correct Markovian long time
asymptotic behavior \cite{pra}.

The different structure of the two Master Equations given by Eq.
\eqref{eq:secIII11} and Eq. \eqref{MeRWA}, traceable back to the
two different coupling Hamiltonians, are responsible for the
occurrence of some physically transparent changes in the
oscillator dynamics, more marked in short time regime.

To better understand the physical origin of such differences let
us have a closer look at the two interaction Hamiltonians:
\begin{subequations}
\begin{equation}\label{RWA}
\hat{H}^{\rm RWA}_{sr} =  \alpha \left(\hat{a} \hat{R}^{\dag} +
 \hat{a}^{\dag} \hat{R}\right)
\end{equation}
 \begin{equation}\label{FV}
 \hat{H}^{\rm FV}_{sr} =  \alpha \left[\left(\hat{a}
\hat{R}^{\dag} +
 \hat{a}^{\dag} \hat{R}\right)+\left(\hat{a}
\hat{R} +
 \hat{a}^{\dag} \hat{R}^{\dag}\right)\right]
 \end{equation}
\end{subequations}

\begin{figure}
\begin{center}
\includegraphics[width=6 cm,height=12 cm, angle=90]{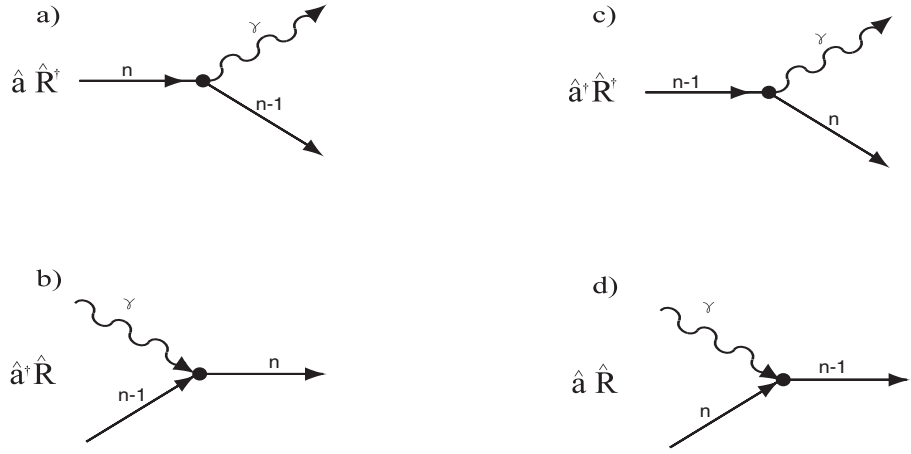}
\caption{\footnotesize There are four distinct terms in the
Hamiltonian given by Eq. \eqref{FV}. The events represented in the
first two diagrams (a,b) correspond to  real processes.The last
two diagrams (c,d), instead, describe events  corresponding to
virtual processes.}\label{diagra}
\end{center}
\end{figure}

We take advantage of a pictorial representation of the four
different interaction terms appearing in Hamiltonian \eqref{FV}
(see fig \ref{diagra}).

\begin{figure}
\begin{center}
\includegraphics[width=4 cm,height=12 cm,
angle=90]{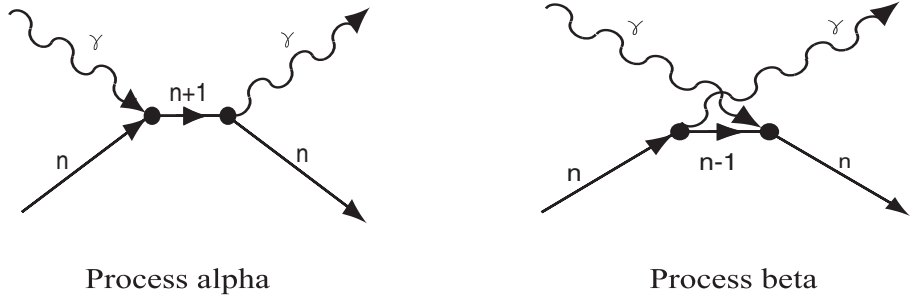} \caption{\footnotesize Two real processes
combine to give rise to the real process alpha. Two virtual
processes combine to give rise to the real process beta.
}\label{2ndorder}
\end{center}
\end{figure}

The events represented in the first two diagrams (a,b) are
processes of absorption or emission in which energy is conserved.
The last two diagrams (c,d), on the contrary, describe events not
corresponding to real absorption and emission processes. For this
reason such processes are called \textit{virtual processes}. In
the second order in perturbation theory both the two real and
virtual processes combine to give rise to  real processes
hereafter called alpha and beta processes respectively (see
fig.\ref{2ndorder}).

Thus when we use the Feynman-Vernon coupling instead of the
Rotating Wave one, the channels through which the oscillator
exchanges energy with the reservoir are doubled. The asymptotic
long time behavior describes, of course, thermalization in both
cases.

These particular features give rise to  different predictions of
the short time behavior of physical quantities, such as for
example
 the mean number of quanta $\med{\hat{n}}(t)$  of the system oscillator (heating function),
 depending on which of the two system-reservoir coupling models is used.
We show in the following that such different behaviors are, in
principle,  experimentally observable and thus relevant for the
correct description of the complete dynamics of the system.

Let us consider, as initial state of the system, the vacuum. It is
well known that, in this case, due to the interaction with the
thermal reservoir at $T$ temperature, the system experiences
heating processes leading to thermalization. In reference
\cite{pra} it has been shown that, for the FV coupling,  the
non-Markovian time evolution of $\med{\hat{n}}(t)$, in the weak
coupling limit, is given by \cite{prl,pra}
\begin{equation}
\med{\hat{n}}(t)\xrightarrow{t<<\omega_c^{-1}}\left[2\alpha^2
\int_0^{\infty} \omega
\abs{g(\omega)}^2(n(\omega)+\frac{1}{2})d\omega \right]
\frac{t^2}{2}, \label{nshort}
\end{equation}
where $n(\omega)$ is the mean number of reservoir excitations at
$T$ temperature and $g(\omega)$ is the reservoir spectral density.

 For the RW coupling, similar calculations yields the
following expression
\begin{equation}
\med{\hat{n}}^{\rm RWA }(t)\xrightarrow{t<<\omega_c^{-1}}\left
[\alpha^2 \int_0^{\infty} \omega \abs{g(\omega)}^2
n(\omega)d\omega\right] \frac{t^2}{2}. \label{nshortRWA}
\end{equation}

Comparing these last two equations one sees immediately that, for
short time intervals, $\med{\hat{n}}(t)\approx 2
\med{\hat{n}}^{\rm RWA }(t)$, meaning that the system--reservoir
FV coupling model predicts an
 initial heating of the system faster than the one
 predicted by the RW coupling model.  This fact can be easily traced back to the
doubling of channels for energy exchange illustrated in Fig. (1).

On the other hand, in the long time asymptotic limit,
$\med{\hat{n}}(t)$ and $\med{\hat{n}}^{\rm RWA}(t)$ have, as
expected, the same temporal behavior \cite{prl,pra}:
\begin{equation}
\med{n}(t>>\omega_0^{-1})=\med{n}^{\rm RWA
}(t>>\omega_0^{-1})\simeq n(\omega_0)(1-exp[-\pi\alpha^2
\omega_0\abs{g(\omega_0)}^2 t]), \label{nasintotico}
\end{equation}
since, due to the time--energy uncertainty principle, for long
times $t$, $\beta$ processes (see Figure \ref{2ndorder}) are very
unlikely to happen in the weak coupling regime.

Summing up the two system--reservoir coupling models under
scrutiny predict the same asymptotic long time behavior for the
observable $\med{\hat{n}}(t)$ but different non--Markovian short
time behaviors. It is worth noting that, once known the system and
reservoir parameters, the only phenomenologic constant is the
coupling constant $\alpha$. Such quantity is usually  estimated
from the experiments \cite{engineer}.
If we now assume that experiments may be performed in all the
relevant time scale, that is both in the asymptotic long time
regime and in the non--Markovian short time regime, one can use
the value of the coupling constant experimentally measured in the
asymptotic long time regime (see Eq. \eqref{nasintotico}) to
verify if the correct short time behavior is actually the one
predicted by Eq. \eqref{nshort} (FV coupling) or the one given by
Eq. \eqref{nshortRWA} (RW coupling). In fact, one would expect
that, since the complete Feynman--Vernon coupling is more general
than the RW coupling, it is also more fundamental and thus it
should give the correct description of the dynamics of the system.

\section{The RWA in the Feynman-Vernon model: comparison with the RW model}

Let us now consider again  the final form of the generalized
Master Equation, given by Eq. \eqref{eq:secIII11} with Eqs.
\eqref{eq:secIII12a}-\eqref{eq:secIII12b}, derived for the FV
coupling. To further simplify the calculation one could think to
perform a Rotating Wave Approximation (RWA) averaging on an
interval $\Delta t$ the rapidly oscillating trigonometric
functions appearing in Eq.(\ref{eq:secIII11}) through
Eqs.\eqref{eq:secIII12a}-\eqref{eq:secIII12b}. Under such
conditions, that is for $2 \omega_0 \Delta t \gg 1$, Eq.
\eqref{eq:secIII11} assumes the form
\begin{equation}
\frac{d \hat{\rho}(t)}{dt} =  -  \left[ \frac{\bar{\Delta}(t)}{2}
\left( (\mathbf{X}^S)^2 + (\mathbf{P}^S)^2 \right) +
i\frac{\gamma(t)}{2} \left(\mathbf{X}^S  \mathbf{P}^{\Sigma}  -
\mathbf{P}^S \mathbf{X}^{\Sigma}\right) \right]\hat{\rho}(t).
\label{eq:secIII13}
\end{equation}
Having in mind Eqs.(\ref{eq:intro3}) and (\ref{eq:secIII7}), after
some straightforward calculations the following Master Equation is
obtained
\begin{eqnarray}
\frac{ d \hat{\rho}}{\partial t}= &-& \frac{\bar{\Delta}(t) +
\gamma (t)}{2} \left[ \hat{a}^{\dag} \hat{a} \hat{\rho} - 2
\hat{a} \hat{\rho} \hat{a}^{\dag} + \hat{\rho} \hat{a}^{\dag}
\hat{a} \right]
\nonumber \\
&-& \frac{\bar{\Delta}(t) - \gamma (t)}{2} \left[  \hat{a}
\hat{a}^{\dag} \hat{\rho} - 2 \hat{a}^{\dag} \hat{\rho} \hat{a} +
\hat{\rho} \hat{a} \hat{a}^{\dag}
 \right], \label{eq:secIII14}
\end{eqnarray}
with $\bar{\Delta}(t)$ and $\gamma(t)$ defined by
Eqs.(\ref{eq:secIII9a})-(\ref{eq:secIII9b}). It is important to
note that Eqs.(\ref{eq:secIII14}) is in the Lindblad form as far
as the sum and difference coefficients $(\bar{\Delta}(t) - \gamma
(t))$ and $(\bar{\Delta}(t) + \gamma (t))$ are positive
\cite{prl}.

\begin{figure}
\includegraphics{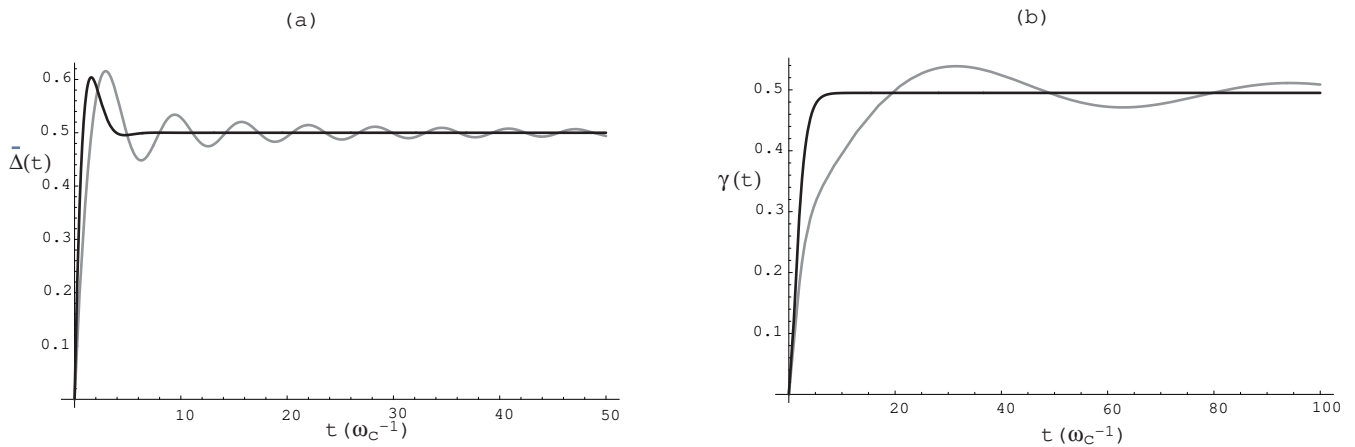}
\caption{\footnotesize (a) Asymptotic long time behavior of the
coefficients $\bar{\Delta}(t)$ (black line) and $\bar{\Delta}^{\rm
RWA}(t)$ (gray line). (b) Asymptotic long time behavior of the
coefficients $\gamma(t)$ (black line) and $\gamma^{\rm RWA}(t)$
(gray line). In both graphics we have put $\omega_c = \omega_0$}
\label{deltas}
\end{figure}

Another interesting feature of Eq. \eqref{eq:secIII14} is that it
has the same structure of the Master Equation obtained starting
from the RW coupling (see Eq. \eqref{MeRWA}). Indeed, one sees
immediately that the difference between the Master Equation
obtained starting form the FV coupling and performing after the
RWA, and that one obtained starting from the RW coupling
\emph{relies only on the time dependent coefficients of the ME}.
Let us have a closer look at the form of such coefficients. In the
limit of continuous modes they are written as:
\begin{subequations}
\begin{equation}
\label{espdelta} \bar{\Delta}(t)=2 \alpha^2
\int_0^t\int_0^{\infty} \omega
\abs{g(\omega)}^2(n(\omega)+\frac{1}{2})\cos[\omega\tau]\cos[\omega_0\tau]d\omega
d\tau ,
\end{equation}
\begin{equation}
\label{espdelta2} \bar{\Delta}^{\rm RWA}(t)=\alpha^2
\int_0^t\int_0^{\infty} \omega
\abs{g(\omega)}^2(n(\omega)+\frac{1}{2})\cos[(\omega-\omega_0)\tau]d\omega
d\tau ,
\end{equation}
\end{subequations}
\begin{subequations}
\begin{equation}
\label{gammfun} \gamma(t)=2\alpha^2 \int_0^t\int_0^{\infty}
\frac{\omega}{2}
\abs{g(\omega)}^2\sin[\omega\tau]\sin[\omega_0\tau]d\omega d\tau,
\end{equation}
\begin{equation}
\label{gammfun2} \gamma^{\rm RWA}(t)=\alpha^2
\int_0^t\int_0^{\infty} \frac{\omega}{2}
\abs{g(\omega)}^2\cos[(\omega-\omega_0)\tau]d\omega d\tau,
\end{equation}
\end{subequations}
where $n(\omega)=\left(exp[\frac{\hbar \omega}{k
T}]-1\right)^{-1}$  is the number of reservoir excitations at $T$
temperature. In the following we assume an Ohmic environment
characterized by a reservoir spectral density having frequency
cut-off $\omega_c$, as for example the Drude spectral density
\begin{equation}
\abs{g(\omega)}^2=\frac{1}{\pi}\frac{\omega_c^2}{\omega_c^2+\omega^2}
\qquad .
\end{equation}

A noticeable difference between the $\bar{\Delta}(t)(\gamma(t))$
and $\bar{\Delta}^{\rm RWA}(t)(\gamma^{\rm RWA}(t))$ coefficients
is that in the last one the anti-resonant term
$\cos[(\omega+\omega_0)\tau]$ is absent. Such a circumstance leads
to distinguishable short time behaviors of the FV and RW
coefficients.

It is indeed possible to prove that in $\bar{\Delta}(t)$, for
$t\ll\omega_c^{-1}$, alpha and beta processes give rise to the
same contributions linear in $t$ so that $\bar{\Delta}(t)\approx
2\bar{\Delta}^{\rm RWA}(t)$.

As far as $\gamma(t)$ is concerned, on the contrary, the same
processes cancel each other at the first order in $t$ in such a
way that, for $t\ll\omega_c^{-1}$, $\gamma(t)\propto t^3$ whereas
$\gamma^{\rm RWA}(t)\propto t$.

In the asymptotic Markovian long time regime we have, as expected,
that $\bar{\Delta}(t \gg \omega_c^{-1}) \simeq \bar{\Delta}^{\rm
RWA }(t \gg \omega_c^{-1}) $ and $\gamma(t \gg \omega_c^{-1})
\simeq \gamma^{\rm RWA }(t \gg \omega_c^{-1}) $, as shown in Fig.
\ref{deltas}

At this point it is worth making some considerations on the
validity of the RWA performed to derive Eq.\eqref{eq:secIII14}. As
we have already said at the beginning of this section, the RWA
consists in neglecting terms oscillating at the frequency $2
\omega_0$. In other words performing the RWA amounts at looking at
the course-grained structure of the dynamics of the systems. For
this reason we cannot describe correctly the dynamical features in
a time interval such as $\Delta t \lesssim \omega_0^{-1}$. Very
often one deals with situations in which the characteristic
frequency of the system $\omega_0$ is smaller or much smaller than
the reservoir frequency cut $\omega_c$. Under this circumstances,
normally, we cannot rely on the short time expressions of the FV
coefficients $\bar{\Delta}(t)$ and $\gamma(t)$ since they are
valid for times $t \ll \omega_c^{-1} \ll \omega_0^{-1} $. However,
there are two cases in which one can use  the Master Equation
given by Eq. \eqref{eq:secIII14} to describe correctly the
non-Markovian short time behavior of the system:
\begin{enumerate}
\item
whenever one wants to look at situations in which $\omega_0
> \omega_c$, as discussed for example in \cite{petruccione,heatingtheory1};
\item
whenever we are interested in the mean value of a certain class of
observables, like for instance the number operator $\langle
\hat{a}^{\dag}\hat{a} \rangle (t)$ (see \cite{prl}).
\end{enumerate}
In this last case, indeed, it has been shown \cite{prl} that,in
the weak coupling limit, it is equivalent to use the solution of
the the Master Equations \eqref{eq:secIII11} or
\eqref{eq:secIII14} since they lead to the the same analytic
expressions for the expectation value of the observable of the
class before mentioned.

\section{Summary and Conclusions}

We have described a procedure, based on  superoperator formalism,
to derive, in the weak coupling limit, non--Markovian generalized
Master Equations local in time. Such a method is equivalent to the
time--convolutionless projection operator technique in the sense
that it leads to the same generalized Master Equation. We apply
this procedure to derive the Master Equation for a specific
system, namely a quantum harmonic oscillator coupled to a thermal
reservoir at $T$ temperature. We compare two different microscopic
system--reservoir coupling models: the Feynman--Vernon and the
rotating wave couplings. Both couplings are bilinear, but the
first one is more general and thus, in this sense, it is more
fundamental. Very often however, in quantum optics systems, the
Rotating Wave coupling is used because the counter rotating terms
not conserving the unperturbed energy cannot contribute to the
system dynamics \cite{cfunction}. The main result of our paper is
to establish under which conditions such a claim is effectively
correct. By comparing the analytic solutions of the heating
function relative to the two different coupling models (FV and RW
couplings) we conclude that, even in the weak coupling limit, the
counter rotating terms give indeed a significant contribution in
the non--Markovian short time regime. Such a contribution is
actually experimentally measurable, provided that one can perform
experiments in all the time scale relevant for the system
dynamics. To this purpose it is worth noting that in the context
of trapped ions experiments have been performed in which the
system (single harmonic oscillator) is first cooled down to its
zero point energy and then coupled to a properly engineered
reservoir \cite{engineer}. We note that, in such experiments, it
is possible not only to choose at will the reservoir parameters,
but also to engineer the coupling and control the coupling
strength. Therefore, the great experimental advances of the
trapped ion techniques could make it possible to perform an
experiment aimed at proving the relevant role, in the short time
dynamics, of the usually neglected counter rotating terms.

One of the reasons for which one usually prefers to work with
master equations derived starting from the RW coupling model is
related to the fact that the resulting Master Equation, in the
Born--Markov approximation, is in the Lindblad form differently
from the case  in which the Feynman--Vernon coupling is assumed
(see Master Equation for Brownian motion). We have demonstrated
here that also the non--Markovian Master Equation obtained
starting from the RW coupling is in the Lindblad form, for some
value of the relevant system and reservoir parameter. Moreover, by
looking at the analytic expression of the time dependent
coefficients of our non--Markovian generalized Master Equations
one can infer the conditions under which one passes from Lindblad
to non Lindblad Master Equations. Remembering that the Master
Equations given by Eqs. \eqref{MeRWA} and \eqref{eq:secIII14} are
of Lindblad type when their time dependent coefficients are
positive, indeed, it is not difficult to convince oneself that
such conditions are simply related to the change of the sign of
the coefficients. Therefore the form of the RW Master Equations
derived in this paper allows to study the border separating two
very different physical regimes characterized by very different
system dynamics \cite{petruccione} and, for this reason, makes it
possible to gain more insight in the fundamental dissipative
processes of one of the most extensively studied physical systems:
a harmonic oscillator coupled to a thermal reservoir.

Another new result we have obtained in this paper stems from  the
comparison between the master equations derived in the following
two cases:

1) Feynman--Vernon system reservoir coupling followed by the RWA
performed after tracing over the reservoir degrees of freedom;

2) Rotating Wave system reservoir coupling.

Stated another way we look at the differences in the system
dynamics arising from the two following approximations
respectively:

1) average over rapidly oscillating terms after tracing over the
reservoir variables;

2) neglecting the counter rotating terms in the initial
microscopic coupling model.

We have shown that the Master Equation obtained from the
Feynman--Vernon coupling, after performing the RWA, is of the
Lindblad type and it actually has the same structure of the RW
Master Equation, with different time dependent coefficients. We
have demonstrated that these two different approximations lead to
different  short time behaviors, while in the asymptotic long time
Markovian regime the two correspondent Master Equations do
coincide. However we have proved that performing the RWA after
tracing over the reservoir variables is a less restrictive
approximation  than starting with the RW coupling model. Indeed,
differently from the RW Master Equation, the Feynman--Vernon one +
RWA, takes into account the virtual photon exchanges relevant in
the short time dynamics and thus it predicts the correct
non--Markovian short time behavior, provided that $t \gg
\omega_0^{-1}$.

\section{Acknowledgements}
One of the authors (S.M.) acknowledges financial support from
Finanziamento Progetto Giovani Ricercatori anno 1999, Comitato 02.

\appendix
\section{The general Master Equation}

In this Appendix we sketch the derivation of Eq. \eqref{eq:secI15}
from Eq. \eqref{eq:secI14}. Let us consider an interaction
Hamiltonian with the form
\begin{equation}
\hat{H}_I (t)= \alpha \hat{A}_i(t)  \hat{E}_i(t)= \alpha
\hat{\vec{A}}(t) \cdot  \hat{\vec{E}}(t),
\end{equation}
where for simplicity we have used Einstein notation. Using some
algebraic properties of the superoperators one can show that if
$[\hat{A}_i,\hat{E}_i]=0$ then
\begin{equation}
(\hat{A}_i(t)
\hat{E}_i(t))^S=\frac{1}{2}\left(\mathbf{A_i}^S\mathbf{E_i}^{\Sigma}+
\mathbf{A_i}^{\Sigma}\mathbf{E_i}^S\right).
\end{equation}

Exploiting the properties of the trace and the assumption
$\med{\hat{E}}=0$ one gets
\begin{equation}
\tr{\mathbf{E}_i^S \hat{\rho}_E(0)}{E}\equiv0, \qquad
\tr{\mathbf{E}_i^{\Sigma} \hat{\rho}_E(0)}{E}=0.
\end{equation}

Consequently
\begin{equation}
\med{\mathbf{H_I}^S(t)}=\tr{\mathbf{E}_i^S
\hat{\rho}_E(0)}{E}=\frac{1}{2}\left(\med{\mathbf{E}_i^{\Sigma}(t)}\mathbf{A}_i^S(t)+
\med{\mathbf{E}_i^S(t)}\mathbf{A}_i^{\Sigma}(t)\right)=0.
\end{equation}

In the same manner it is not difficult to show that
\begin{equation}
\med{\mathbf{H_I}^S(t)\mathbf{H_I}^S(t_1)}=\frac{1}{4}\left(
\med{\mathbf{E}_i^{\Sigma}(t)\mathbf{E}_j^{\Sigma}(t_1)}\mathbf{A}_i^S(t)\mathbf{A}_j^S(t_1)+
\med{\mathbf{E}_i^{\Sigma}(t)\mathbf{E}_j^S(t_1)}\mathbf{A}_i^S(t)\mathbf{A}_j^{\Sigma}(t_1)\right),
\end{equation}
where the equalities
\begin{equation}
\med{\mathbf{E}_i^S(t)\mathbf{E}_j^{\Sigma}(t_1)}=
\med{\mathbf{E}_i^S(t)\mathbf{E}_j^S(t_1)}\equiv0,
\end{equation}
have been used. After some algebraic manipulation one gets
\begin{gather}
\med{\mathbf{E}_i^{\Sigma}(t)\mathbf{E}_j^{\Sigma}(t_1)}= 2\med{\{
\hat{E}_i^{\Sigma}(t-t_1),\hat{E}_j^{\Sigma}(0)
\}},\\
\med{\mathbf{E}_i^{\Sigma}(t)\mathbf{E}_j^S(t_1)}=
2\med{[\hat{E}_i^{\Sigma}(t-t_1),\hat{E}_j^{\Sigma}(0) ]},
\end{gather}
where the square an curl brackets indicate the commutator and
anti-commutator respectively.

Substituting these expressions into Eq.\eqref{eq:secI14} and using
the definitions of the correlation and susceptibility matrices one
obtains the general Master Equation given by \eqref{eq:secI15}.

\section{Derivation of FV Master Equation}

In this appendix we present the superoperatorial mathematical
properties allowing to derive the final form of the FV Master
Equation, given by Eq. \eqref{eq:secIII11} discussed in this
paper. First of all let us consider the following superoperatorial
relations
\begin{subequations}
\label{comm}
\begin{gather}
[\mathbf{A}^S,\mathbf{B}^S]=[\mathbf{A}^{\Sigma},\mathbf{B}^{\Sigma}]=[\hat{A},\hat{B}]^S\\
[\mathbf{A}^S,\mathbf{B}^{\Sigma}]=[\hat{A},\hat{B}]^{\Sigma}.
\end{gather}
\end{subequations}

Such these equations and having in mind the Baker-Hausdorff
formula one gets
\begin{equation}
\mathbf{A}^{S(\Sigma)}(t)\equiv\exp \left[ \imath \mathbf{B}^{S} t
\right]\:\mathbf{A}^{S(\Sigma)}\:\exp \left[ -\imath
\mathbf{B}^{S} t \right]=\left( \exp \left[ \imath \hat{B}^{S}
 t \right]\:\hat{A} \:\exp \left[ -\imath \hat{B}^{S} t
\right]\right)^{S(\Sigma)}=\left(\hat{A}(t)\right)^{S(\Sigma)}.
\end{equation}
The previous relation says that the time evolution of
superoperators is ruled by equations analogue to those of the
operators in Dirac's formalism. Then, specifying Eq.
\eqref{eq:secI15} to the system under scrutiny and using Eqs.
\eqref{eq:secIII1} and \eqref{eq:secIII5} yields
\begin{subequations}
\begin{eqnarray}
\mathbf{D}_S(t) &=&\int_0^t \kappa (\tau)
\mathbf{T}_0(t)\:\mathbf{X}^S(t)\mathbf{X}^S(t-\tau)\:\mathbf{T}^{-1}_0(t)d\tau=\int_0^t
\kappa (\tau)
\mathbf{X}^S\mathbf{X}^S(-\tau)d\tau, \nonumber\\
&=& \int_0^t\kappa (\tau) \mathbf{X}^S \left( \cos{\omega_0 \tau }
\mathbf{X}^S - \sin{\omega_0 \tau } \mathbf{P}^S \right)d\tau
\end{eqnarray}

\begin{eqnarray}
\mathbf{G}_S(t)&=&\int_0^t \mu (\tau)
\mathbf{T}_0(t)\:\mathbf{X}^S(t)\mathbf{X}^{\Sigma}(t-\tau)\:\mathbf{T}^{-1}_0(t)d\tau=\int_0^t
\mu (\tau)
\mathbf{X}^S\mathbf{X}^{\Sigma}(-\tau)d\tau\nonumber\\
&=& \int_0^t \mu (\tau) \mathbf{X}^{S} \left( \cos{\omega_0 \tau }
\mathbf{X}^{\Sigma} - \sin{\omega_0 \tau } \mathbf{P}^{\Sigma}
\right)d\tau.
\end{eqnarray}
\end{subequations}

\section{Derivation of the RW Master Equation}
In this Appendix we underline the essential steps in the
derivation of the RW Master Equation given by Eq.
\eqref{eq:secIV1}. Let us note that, in interaction picture, the
Hamiltonian given by Eq. \eqref{eq:intro1} assumes the form
\begin{equation}
\hat{H}^{\rm RWA}_{sr}(t) = \alpha \sum_{i=0}^{\infty}\hbar
\sqrt{\frac{\omega}{2}}\left( g_i \hat{a}e^{-\imath \omega_0 t}
\hat{b}_i^{\dag}e^{\imath \omega_i t} + h.c. \right).
\end{equation}

From a mathematical point of view it is convenient to associate
all the time dependent phase factors to the reservoir operators as
follows

\begin{equation}
\hat{H}^{\rm RWA}_{sr}(t)\equiv \alpha \left(\hat{a}
\hat{\tilde{R}}^{\dag}(t) +
 \hat{a}^{\dag} \hat{\tilde{R}}(t)\right),
\end{equation}
with
\begin{equation}
\hat{\tilde{R}}(t)\equiv \sum_{i=0}^{\infty}
\hbar\sqrt{\frac{\omega}{2}} g_i \hat{b}_i e^{-\imath(
\omega_i-\omega_0) t}
\end{equation}

 For the system here considered, thus, the operators appearing in the bilinear form defined in Eq.
 \eqref{eq:secI13} are $\hat{\vec{A}}(t)=\hat{\vec{A}}=(\hat{a},\hat{a}^{\dag})$ and
 $\hat{\vec{E}}(t)=(\hat{\tilde{R}}^{\dag}(t),\hat{\tilde{R}}(t))$.

Exploiting the properties of superoperators given by Eq.
\eqref{comm}, with some algebraic manipulation, the Master
Equation given by Eq.\eqref{eq:secI15} can be recast in the form
given by Eq. \eqref{eq:secIV1}. Finally we write such a Master
Equation in the Schr\"{o}dinger picture exploiting of the
following property
\begin{equation}
\mathbf{T}_0(t)\:\mathbf{a^{\dag}}^S\mathbf{a}^{S(\Sigma)}\:\mathbf{T}^{-1}_0(t)=
(\hat{a}^{\dag}\:e^{-\imath\omega_0
t})^S(\hat{a}\:e^{\imath\omega_0
t})^{S(\Sigma)}=\mathbf{a^{\dag}}^S\mathbf{a}^{S(\Sigma)},
\end{equation}
with $\mathbf{T}_0(t)$ defined by Eq. \eqref{eq:secIII3}.
Concluding, we note that, the superoperator proportional to
$r^{\rm RWA}(t)$, appearing in Eq.\eqref{eq:secI15}, can be recast
in the form
\begin{equation}
\frac{1}{2}\left(\mathbf{a^{\dag}}^S\mathbf{a}^{\Sigma}+\mathbf{a}^S\mathbf{a^{\dag}}^{\Sigma}\right)=\left(\hat{a}^{\dag}\hat{a}\right)^S.
\end{equation}
Thus the corresponding term in the RW Master Equation is a
frequency renormalization term. Neglecting this term and going
back to the interaction picture one gets the final form of the
Master Equation, given by Eq. \eqref{MeRWA}.

\end{document}